\documentclass[journal]{IEEEtran}
\usepackage[dvips]{graphicx}
\usepackage{epsfig}
\usepackage{amsmath}
\usepackage{epsfig}
\usepackage{array}
\usepackage{amssymb}
\usepackage{color}

\begin{document}

\title{STEEP -- An Alternative To Quantum Key Distribution}

\author{Yingbo Hua, \emph{Fellow, IEEE}  \thanks{Yingbo Hua is with Department of Electrical and Computer Engineering,
University of California at Riverside, Riverside, CA 92521, USA. Email: yhua@ece.ucr.edu. This work was supported in part by the Department of Defense under W911NF-20-2-0267. The views and conclusions contained in this
document are those of the author and should not be interpreted as representing the official policies, either
expressed or implied, of  the U.S. Government. The U.S. Government is
authorized to reproduce and distribute reprints for Government purposes notwithstanding any copyright
notation herein.}}

\maketitle

\begin{abstract}
Secret-message transmission by echoing encrypted probes (STEEP) is discussed as an alternative to quantum key distribution (QKD). The former only needs classic or non-quantum channels while the latter needs both quantum and classic channels for secret-key generation. STEEP is shown to yield a secrecy rate sufficient for one-time pads encryption in many practical situations including in-air channels or undersea optical cables. Other advantages of STEEP over QKD include cost, complexity, compatibility, and robustness against constant eavesdropping.

\end{abstract}

\section{Introduction}

Quantum Key Distribution (QKD) has attracted enormous interests from both academia and industry, and this trend is projected to grow rapidly in the coming years \cite{QKD_2025}. The major promise of QKD is that two legitimate parties or nodes (Alice and Bob) can establish a secret key by using a quantum channel as well as an authenticated classic channel between themselves. The secrecy of the key is based on the theory that any eavesdropping on the quantum channel is detectable by the legitimate parties. By discarding all potentially eavesdropped quantum signals, QKD can ensure the complete secrecy of its generated key.

However, as pointed out in \cite{QKD_2025} and many references therein, there are critical issues with QKD. The first is the cost, including the costly establishment of a quantum channel along with a classic channel between every two nodes of interest. A pair of quantum transmitter and receiver is by itself much more costly than a transceiver for common optical and/or radio channels. The second issue is the (highly) limited  key rate of QKD over a long distance in comparison to the data rate of a classic channel over the same distance such as undersea optical cable. Quantum repeaters are not like common optical repeaters, the latter of which are very much mature and widely deployed in practice. The third issue is that QKD could be completely paralized  if the quantum channel is subject to constant eavesdropping. In this case, no key with a good confidence could be generated.

Given the limitations of QKD, Post-Quantum-Cryptography (PQC) \cite{PQC_2024} remains the primary defense against future quantum computing powers  to keep the Internet safe from crypto attacks. Yet, unlike QKD, PQC is based on the current understanding of the computational hardness to invert some selected crypto algorithms. If the future computing powers and/or algorithmic discoveries become far more powerful than the current predictions, then the current PQC-based strategy could fail to beat ``Harvest-Now-Decrypt-Later (HNDL)'' attack, and a continued revamping of PQC in the future could be too costly in terms of re-standardizations and re-implementations.

A good news that this paper aims to report is that there is a simple and effective alternative to QKD in providing information-theoretically-secure secret keys between virtually any pair of nodes connected by a classic channel. Such a method is called ``Secret-message Transmission by Echoing Encrypted Probes (STEEP)'' \cite{STEEP_2025}. While many of the theoretical details of STEEP along with its connections to some prior results such as in \cite{Maurer1993}, \cite{Bloch2011}, \cite{Gamal2011}, \cite{Hayashi2020}, \cite{Hua_2023} and \cite{HuaMaksud2024} are shown in \cite{STEEP_2025}, this paper presents STEEP in its simplest form in order to highlight its practical values especially in contrast to QKD. The discussions of STEEP shown in this paper mostly contain improvements and/or new insights not published elsewhere.

\section{STEEP and Its Secrecy Rate}

STEEP is a scheme designed for classic channels.
The simplest classic channel is a single-input-single-output (SISO) additive-white-Gaussian-noise (AWGN) channel, which applies to many radio, optical, coaxial and copper channels in practice. This is the channel that this paper is focused on.

\subsection{The STEEP Scheme}

STEEP considered in this paper is a round-trip transmission scheme between two nodes (Alice and Bob) with two phases.

In phase 1 (P1), Alice  transmits a packet containing uncoded complex random symbols $x_{1,k}$ (also called probes) so that Bob  receives
\begin{equation}\label{eq:y1k}
  y_{1,k}=x_{1,k}+w_{1,k}
\end{equation}
for $k= 1,\cdots,K$.
Here without loss of generality (w.l.o.g.), the channel gain has been normalized to be one since we assume that  the packet from a transmitting node has a sufficiently strong pilot so that the receiving node can obtain an accurate estimate of its receive channel response and perform coherent processing. Furthermore, we can assume that the signal $x_{1,k}$ has the unit variance and the white circular Gaussian (w.c.g.) noise $w_{1,k}$ has the variance $\sigma_1^2$. The normalized transmission power for P1 can be denoted by $p_1=\frac{1}{\sigma_1^2}$, which is also the SNR of the P1 channel shown by \eqref{eq:y1k}.

Note that \eqref{eq:y1k} can be also viewed as one of the subcarrier channels in an OFDM system.
Also the roles of Alice and Bob can be reversed in another STEEP session as described later in section \ref{sec:special_case}.

 In phase 2 (P2) of STEEP, Bob transmits a packet containing such symbols $x_{2,k}$ defined as
\begin{equation}\label{eq:x2k}
  x_{2,k}=c_1 y_{1,k}+c_2 s_{2,k},
\end{equation}
so that Alice receives
\begin{equation}\label{eq:y2k}
  y_{2,k}=x_{2,k}+w_{2,k}
\end{equation}
for $k=1,\cdots,K$. Here the w.c.g. noise $w_{2,k}$ has the variance $\sigma_2^2$, and the signal $x_{2,k}$ with the weights $c_1$ and $c_2$ is also chosen w.l.o.g. to have the unit variance.
Then the (normalized) transmission power in P2, or the SNR of the P2 channel represented by \eqref{eq:y2k}, is $p_2=\frac{1}{\sigma_2^2}$.

Furthermore,  $s_{2,k}$ for $k=1,\cdots,K$ is a sequence of coded complex symbols (unit variance) so that the message represented by this sequence can be decoded reliably by Alice using her knowledge of both $y_{2,k}$ and $x_{1,k}$ for $k=1,\cdots,K$.

It follows that
\begin{equation}\label{eq:c1c2}
  \mathbb{E}\{|x_{2,k}|^2\}=c_1^2(1+\sigma_1^2)+c_2^2 =1.
\end{equation}
Clearly, w.l.o.g., we can treat $c_1$ and $c_2$ as being non-negative and real-valued.

Here the construction of $x_{2,k}$ shown in \eqref{eq:x2k} allows some freedoms in $c_1$ and $c_2$ subject to \eqref{eq:c1c2}. This is more general than the (unweighted) sum of $s_{2,k}$ and the MMSE estimate of $x_{1,k}$, the latter of which was adopted in \cite{STEEP_2025} for the SISO case.

For Alice to detect the information in $s_{2,k}$ from $y_{2,k}$ and $x_{1,k}$, it is equivalent for her to detect the information in $s_{2,k}$ from
\begin{align}\label{eq:Dy2k}
 & \Delta y_{2,k}\doteq y_{2,k}-\mathbb{E}\{y_{2,k}|x_{1,k}\}=y_{2,k}-c_1x_{1,k}
 \notag\\
 &
  =c_2s_{2,k}+w_{2,k}-c_1w_{1,k}
\end{align}
 for $k=1,\cdots,K$. Here we have applied the independence between $x_{1,k}$ and $s_{2,k}$.

In order to establish the secrecy rate of STEEP, we must consider the signals received by Eve in P1 and P2, which can be written as
\begin{equation}\label{eq:z1k}
  z_{1,k}=x_{1,k}+v_{1,k},
\end{equation}
\begin{equation}\label{eq:z2k}
  z_{2,k}=x_{2,k}+v_{2,k},
\end{equation}
where $v_{1,k}$ and $v_{2,k}$ are w.c.g. with the variances $\epsilon_1^2$ and $\epsilon_2^2$. Also define
$\alpha_1=\frac{\sigma_1^2}{\epsilon_1^2}$ and $\alpha_2=\frac{\sigma_2^2}{\epsilon_2^2}$. Here $\alpha_1$ represents an advantage of Eve's P1 receive channel over user's P1 receive channel, and $\alpha_2$  an advantage of Eve's P2 receive channel over user's P2 receive channel.

In order for STEEP to be competitive against QKD, we must focus on the situations where $\alpha_1>1$ and $\alpha_2>1$. Furthermore, we assume that Eve is constantly present. A simple derivation of the secrecy rate of STEEP is shown next.

\subsection{Secrecy Rate of STEEP}

An achievable secrecy rate from user's transmitter to user's receiver in P2 via $s_{2,k}$ (in bits per round-trip sample interval) is $R_s^+=\max(0,R_s)$ with
\begin{equation}\label{}
  R_s = C_U-C_E
\end{equation}
where $C_U\doteq I(s_{2,k};x_{1,k},y_{2,k}) = I(s_{2,k};y_{2,k}|x_{1,k})$ which is the \emph{effective} channel capacity from Bob to Alice via $s_{2,k}$, and $C_E\doteq I(s_{2,k};z_{1,k},z_{2,k})$ which is the \emph{effective} (maximum possible) leakage capacity from both users to Eve about $s_{2,k}$.

Assuming that all signals and noises are zero-mean w.c.g., we can further write $C_U=- \log_2\sigma_{s_{2,k}|x_{1,k},y_{2,k}}^2$ and $C_E=-\log_2\sigma_{s_{2,k}|z_{1,k},z_{2,k}}^2$ \cite{STEEP_2025}, and hence
\begin{equation}\label{eq:Rs}
  R_s
  =\log_2\sigma_{s_{2,k}|z_{1,k},z_{2,k}}^2 - \log_2\sigma_{s_{2,k}|x_{1,k},y_{2,k}}^2
\end{equation}
where $\sigma_{s_{2,k}|x_{1,k},y_{2,k}}^2$ is the MSE of the MMSE estimate of $s_{2,k}$ by Alice from $x_{1,k}$ and $y_{2,k}$, and $\sigma_{s_{2,k}|z_{1,k},z_{2,k}}^2$ is the MSE of the MMSE estimate of $s_{2,k}$ by Eve from $z_{1,k}$ and $z_{2,k}$. These MSEs are presented next.
\subsubsection{MSE at Alice}
 Recall \eqref{eq:Dy2k}. It follows that
\begin{align}\label{eq:sigma_s2|x1_y2}
  &\texttt{MSE}_U\doteq\sigma_{s_{2,k}|x_{1,k},y_{2,k}}^2\notag\\
  &=1-|\mathbb{E}\{s_{2,k}(\Delta y_{2,k})^*\}|^2(\mathbb{E}\{|\Delta y_{2,k}|^2\})^{-1}\notag\\
  &=1-\frac{c_2^2}{c_2^2+c_1^2\sigma_1^2+\sigma_2^2}
  =\frac{1}{1+\frac{c_2^2}{c_1^2\sigma_1^2+\sigma_2^2}}.
\end{align}
\subsubsection{MSE at Eve}
Now let $\mathbf{z}_k=[z_{1,k},z_{2,k}]^T$. It follows that
\begin{equation}\label{}
  \mathbf{z}_k=\left [\begin{array}{c}
                      x_{1,k}+v_{1,k} \\
                      c_2s_{2,k}+c_1x_{1,k}+c_1w_{1,k}+v_{2,k}
                    \end{array}
  \right ].
\end{equation}
Then
\begin{align}\label{}
  &\sigma_{s_{2,k}|z_{1,k},z_{2,k}}^2
  =1-\mathbb{E}\{s_{2,k}\mathbf{z}_k^H\}(\mathbb{E}\{\mathbf{z}_k\mathbf{z}_k^H\})^{-1}
  (\mathbb{E}\{s_{2,k}\mathbf{z}_k^H\})^H\notag\\
  &=1-\mathbf{a}_k^H\mathbf{A}_k\mathbf{a}_k,
\end{align}
with
\begin{equation}\label{}
  \mathbf{a}_k^H\doteq\mathbb{E}\{s_{2,k}\mathbf{z}_k^H\}
  =[0,c_2],
\end{equation}
\begin{align}\label{}
  &\mathbf{A}_k\doteq \mathbb{E}\{\mathbf{z}_k\mathbf{z}_k^H\}
  =\left [\begin{array}{cc}
            1+\epsilon_1^2 & c_1 \\
            c_1 & c_2^2+c_1^2+c_1^2\sigma_1^2+\epsilon_2^2
          \end{array}
   \right ]\notag\\
   &=\left [\begin{array}{cc}
            1+\epsilon_1^2 & c_1 \\
            c_1 & 1+\epsilon_2^2
          \end{array}
   \right ],
\end{align}
where \eqref{eq:c1c2} has been applied. Since
\begin{equation}\label{}
  \mathbf{A}_k^{-1}
  =\frac{1}{(1+\epsilon_1^2)(1+\epsilon_2^2)-c_1^2}\left [\begin{array}{cc}
            1+\epsilon_2^2 & -c_1 \\
            -c_1 & 1+\epsilon_1^2
          \end{array}
   \right ],
\end{equation}
then
\begin{align}\label{eq:sigma_s2|z1_z2}
  &\texttt{MSE}_E\doteq\sigma_{s_{2,k}|z_{1,k},z_{2,k}}^2
  =1-\frac{c_2^2(1+\epsilon_1^2)}{(1+\epsilon_1^2)(1+\epsilon_2^2)-c_1^2}
  \notag\\
  &=\frac{1}{1+\frac{c_2^2(1+\epsilon_1^2)}{(1+\epsilon_1^2)(1+\epsilon_2^2-c_2^2)-c_1^2}}.
\end{align}

\subsubsection{Secrecy rate of STEEP}
Using \eqref{eq:sigma_s2|x1_y2} and \eqref{eq:sigma_s2|z1_z2} in \eqref{eq:Rs} yields
\begin{equation}\label{eq:Rs_2}
  R_s = \log_2\frac{1+\frac{c_2^2}{c_1^2\sigma_1^2+\sigma_2^2}}
  {1+\frac{c_2^2(1+\epsilon_1^2)}{(1+\epsilon_1^2)(1+\epsilon_2^2-c_2^2)-c_1^2}}.
\end{equation}
Recall $\sigma_1^2=\frac{1}{p_1}$, $\sigma_2^2=\frac{1}{p_2}$, $\epsilon_1^2=\frac{1}{\alpha_1p_1}$, $\epsilon_2^2=\frac{1}{\alpha_2p_2}$ and the constraint \eqref{eq:c1c2} on $c_1$ and $c_2$.
%
Hence $R_s$ is determined by $p_1, p_2, \alpha_1, \alpha_2, c_1^2$ subject to $0<c_1^2<\frac{1}{1+\frac{1}{p_1}}$.

\section{General Properties of STEEP's Secrecy Rate}

One can verify the following properties of $R_s$ shown in \eqref{eq:Rs_2}:

1) $R_s$ is a decreasing function of  $\alpha_1$ and $\alpha_2$. This is expected intuitively since $\alpha_1$ and $\alpha_2$ increase as Eve's receive channel strength increases in P1 and P2 respectively.

2) For $\alpha_2\geq 1$, $R_s$ is an increasing function of $p_2$. This is also expected intuitively because Eve's channel advantage in P2 (when $\alpha_2> 1$) becomes less and less important as $p_2$ increases.

3) As $p_1\to\infty$ and $\frac{p_2}{p_1}\to\infty$, $R_s$ always converges to a positive value. Specifically,
  \begin{align}
&\lim_{p_2\to\infty} R_s=\log_2\frac{1+\frac{1- c_1^2\left (1+\frac{1}{p_1}\right)}{c_1^2\frac{1}{p_1}}}
  {1+\frac{\left (1- c_1^2\left (1+\frac{1}{p_1}\right)\right )\left (1+\frac{1}{\alpha_1p_1}\right )}{\left (1+\frac{1}{\alpha_1p_1}\right)
   c_1^2\left (1+\frac{1}{p_1}\right)-c_1^2}}\notag\\
   &=\log_2\left (\frac{1-c_1^2}{1+\frac{1}{\alpha_1p_1}-c_1^2}\left (1+\frac{1}{\alpha_1}+\frac{1}{\alpha_1p_1}\right )\right )
\end{align}
and hence
\begin{equation}\label{eq:Rs_lim}
  \lim_{p_1\to\infty}\lim_{p_2\to\infty} R_s = \log_2\left (1+\frac{1}{\alpha_1}\right ).
\end{equation}
This means that if $p_2\gg p_1\gg 1$, then $R_s$ is virtually invariant to $c_1$ and $c_2$ subject to \eqref{eq:c1c2}. Note that (50) in \cite{STEEP_2025} is a special case of  \eqref{eq:Rs_lim}.

4) The secrecy rate is positive, i.e., $R_s>0$, if and only if
\begin{equation}\label{eq:cond_p2}
  p_2> \left (1-\frac{1}{\alpha_2}\right )\frac{1+\alpha_1p_1}{c_1^2}.
\end{equation}
It follows that there is a $c_1^2<\frac{1}{1+\frac{1}{p_1}}$ to make $R_s>0$ if and only if
\begin{equation}\label{eq:cond_p1_p2_a1_a2}
  p_2>\left (1-\frac{1}{\alpha_2}\right )\left (1+\alpha_1+\alpha_1p_1+\frac{1}{p_1}\right ).
\end{equation}
It also follows from \eqref{eq:cond_p1_p2_a1_a2} that if $p_1\gg 1$, then $p_2>\alpha_1p_1$ is almost surely sufficient to make $R_s>0$.

\begin{figure}
  \centering
  \includegraphics[width=3in]{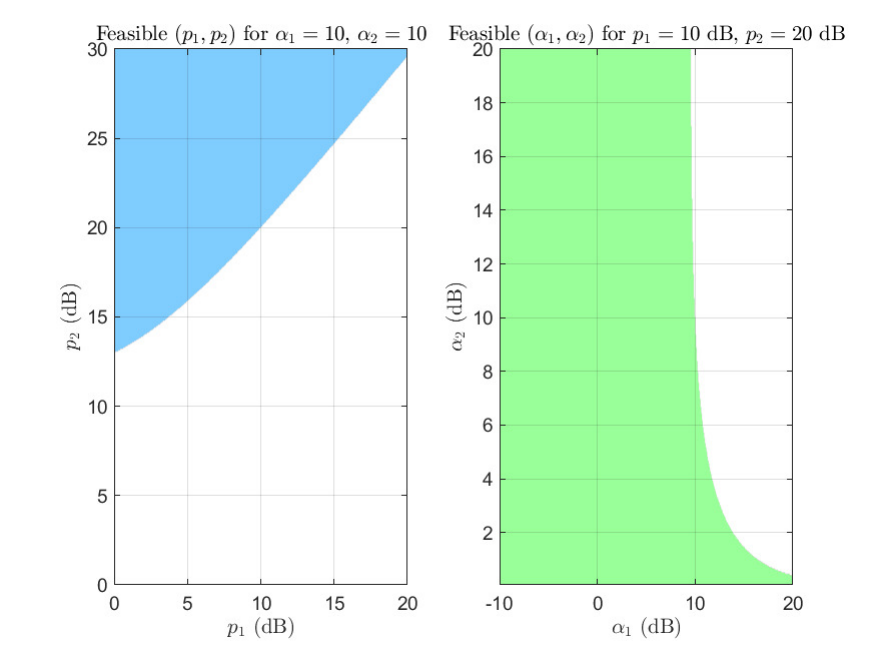}
  \caption{Illustration of the feasible regions governed by \eqref{eq:cond_p1_p2_a1_a2} for $R_s>0$.}\label{Fig:Feasible_regions}
\end{figure}

5) The space of the parameters satisfying \eqref{eq:cond_p1_p2_a1_a2} can be called the overall feasible region for a positive secrecy rate, which is a four-dimensional space. But
 \eqref{eq:cond_p1_p2_a1_a2} also defines a feasible region of $p_1$ and $p_2$  when  $\alpha_1$ and $\alpha_2$ are given, and a feasible region of $\alpha_1$ and $\alpha_2$  when $p_1$ and $p_2$ are given. These two feasible regions are illustrated in Fig. \ref{Fig:Feasible_regions}

\section{Special Cases of STEEP's Secrecy Rate}\label{sec:special_case}

\begin{figure}
  \centering
  \includegraphics[width=3in]{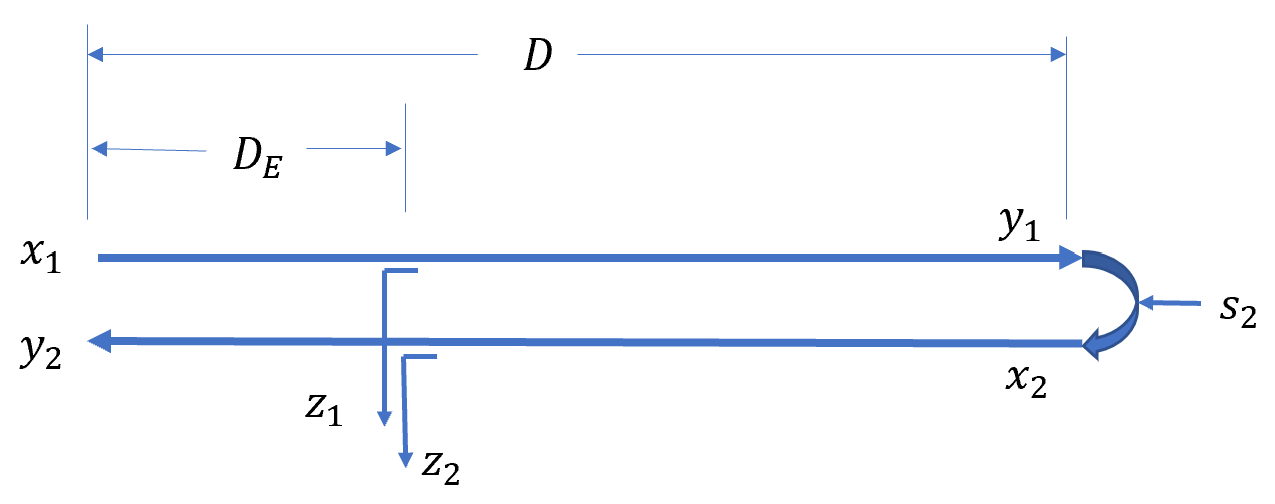}
  \caption{A pair of ``forward and backward''  wireline channels, or ``downlink and uplink'' or ``uplink and downlink'' wireless channels, over which STEEP is applied.}\label{Fig:Forward_backward_channels}
\end{figure}

We now consider the case where Eve is located in  between Alice and Bob, which is the worst case. See Fig. \ref{Fig:Forward_backward_channels}.
Let the distance between Alice and Bob be $D$, and the distance between Alice and Eve be $D_E$. Also define
\begin{equation}\label{}
  d\doteq\frac{D_E}{D}.
\end{equation}

For a pure wireline channel such as optical, coaxial or copper cable, the power loss of signal propagation is an exponential function of the length of the cable. But for a long-haul wireline channel such as an undersea optical cable covering hundreds to thousands miles, repeaters (one per 80 km, for example) are necessary to maintain a sufficient SNR at the destination. For many reasons (such as reduced cost and complexity, wavelength-division-multiplexing  compatibility, reduced power consumption and heat dissipation, and reliability and maintenance), modern optical repeaters are non-regenerative, which makes the SNR at the destination scale with $D$ like $\frac{1}{D^\alpha}$ with $1\leq \alpha\leq 2$. This scaling law accounts for amplified spontaneous emission, nonlinear interference and spectral broadening and distortion. In principle, this scaling law also applies to all long-haul wireline channels (including coaxial and copper cables) using non-regenerative repeaters \cite{Agrawal_2021} and \cite{Pozar_2005}. For line-of-sight wireless channels such as most satellite communication channels, the path loss exponent is fixed at $\alpha=2$ \cite{Tse_2005}.

So, for a long-haul wireline channel or a line-of-sight wireless channel,  we can represent Eve's SNR in P1 by $\eta d^{-\alpha} p_1$, and Eve's SNR in P2 by $\eta (1-d)^{-\alpha}p_2$. Here  $\eta$ represents the efficiency of the wiretapper (Eve). If Eve's receiver is as efficient as the legitimate receiver, we have $\eta=1$. However, for realistic scenarios where Eve suffers many practical hurdles, she (with single antenna or sensor as legitimate receiver) can only achieve $\eta<1$ if not $\eta\ll 1$.

In this case, the channel advantages of Eve in P1 and P2 are $\alpha_1=\eta d^{-\alpha}$ and $\alpha_2=\eta(1-d)^{-\alpha}$, respectively. To be conservative against the worst cases, we will choose $\alpha=2$ and $\eta=1$. For a realistic SNR at legitimate receiver, we will also choose $p_2=20$dB.

\begin{figure}
  \centering
  \includegraphics[width=3in]{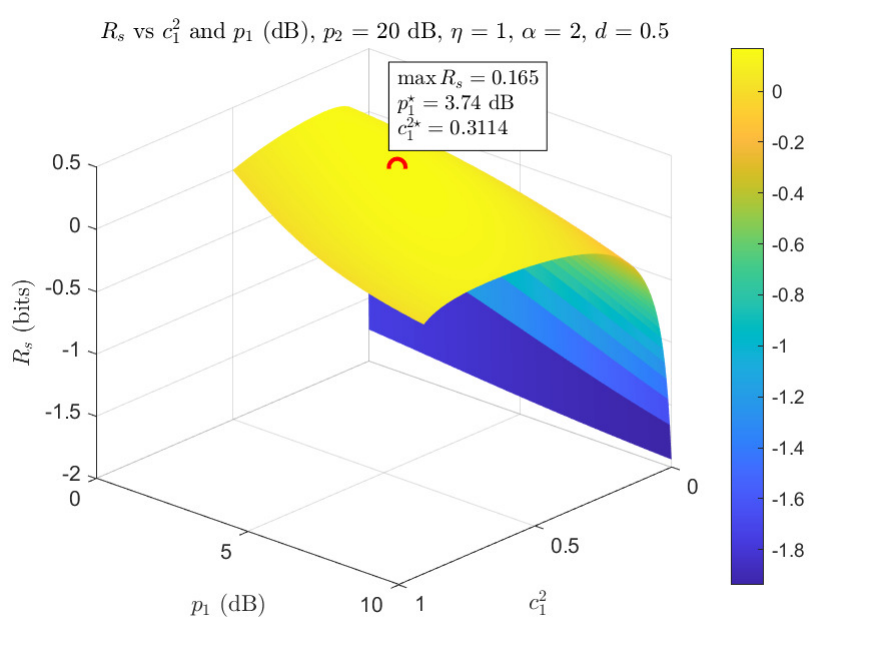}
  \caption{$R_s$ versus $p_1$ and $c_1^2$ with $p_2=20$dB.}\label{Fig:Rs_vs_p1_c1_sq}
\end{figure}

 Fig. \ref{Fig:Rs_vs_p1_c1_sq} illustrates how $R_s$ with $d=0.5$ varies with $p_1$ and $c_1^2$, which shows that the peak value of $R_s$ is relatively insensitive to $p_1$ and $c_1^2$.

\begin{figure}
  \centering
  \includegraphics[width=3in]{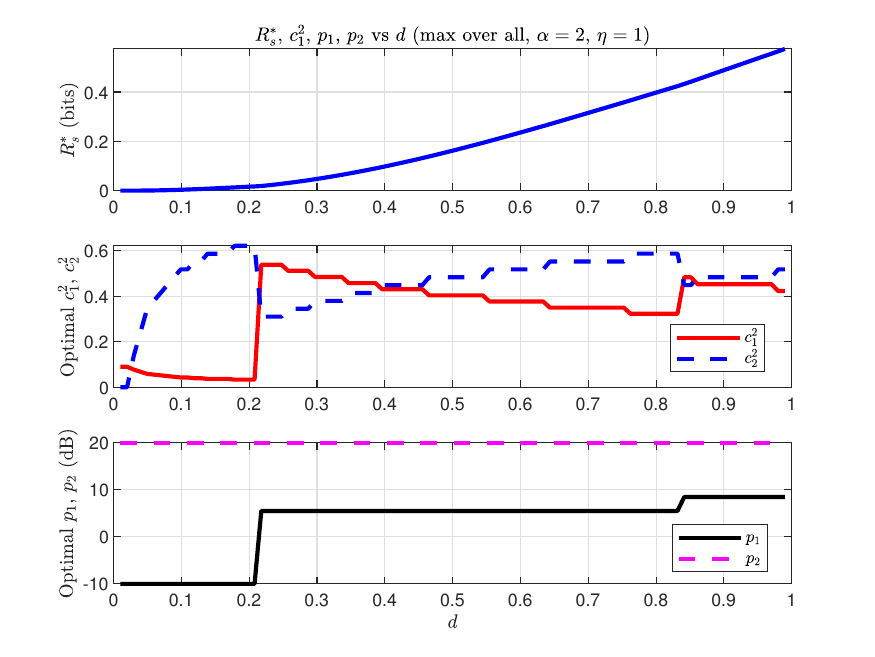}
  \caption{$R_s^*\doteq \max_{p_1,c_1}R_s$ and the joint optimal $p_1$ and $c_1$ versus $d$}\label{Fig:optimal_Rs_over_c1_and_p1_vesus_d}
\end{figure}

Now let
\begin{equation}\label{}
  R_s^*\doteq\max_{p_1,c_1}R_s,
\end{equation}
and $(p_1^*,c_1^*)=arg\max_{p_1,c_1}R_s$. Note that $p_1^*$  can be easily realized in practice by controlling the transmit power in P1.

 Fig. \ref{Fig:optimal_Rs_over_c1_and_p1_vesus_d} illustrates how $R_s^*,c_1^*, p_1^*$ vary with $d$. In this case, both $\alpha_1$ and $\alpha_2$ are larger than one for all $0<d<1$. For reasons shown in Fig. \ref{Fig:Rs_vs_p1_c1_sq},  the jointly optimal $p_1$ and $c_1^2$ do not change much over a wide range of $d$ (e.g., for $0.25<d<0.8$ here).

\begin{figure}
  \centering
  \includegraphics[width=3in]{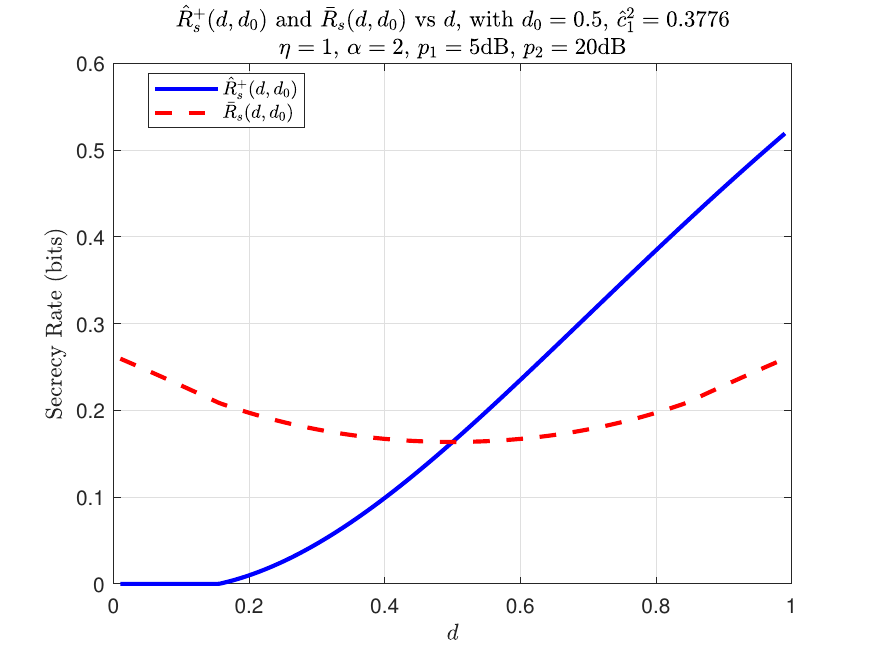}
  \caption{$\hat R_s^+(d_0,d)$ and $\bar R_s(d_0,d)$ with $d_0=0.5$ versus $d$}\label{Fig:Rs_optimized_at_d=0.5_vs_d}
\end{figure}

Let
$\hat R_s\doteq\max_{c_1}R_s$
and $\hat c_1
\doteq arg\max_{c_1}R_s$. If we choose $\hat c_1$ computed at $d=d_0$ and use it for all $0<d<1$, the resulting $R_s$ is now denoted by  $\hat R_s(d_0,d)$.

Also define an averaged secrecy rate (still in bits per round-trip sample):
\begin{equation}\label{}
  \bar R_s(d_0,d)\doteq\frac{1}{2}\left [\hat R_s^+(d_0,d) +\hat R_s^+(d_0,1-d)\right ],
\end{equation}
  where $\hat R_s(d_0,1-d)$ is the effective secrecy rate from Alice to Bob against the same Eve (positioned at the same distance from Alice) when the transmitting nodes in P1 and P2 are assigned to Bob and Alice respectively (instead of Alice and Bob). For both $\hat R_s(d_0,d)$ and $\hat R_s(d_0,1-d)$, the same $c_1$ (along with the same $p_1$ for P1 transmitter and the same $p_2$ for P2 transmitter) is used.

Fig. \ref{Fig:Rs_optimized_at_d=0.5_vs_d} illustrates $\hat R_s^+(d_0,d)$ and $\bar R_s(d_0,d)$ versus $d$ with $d_0=0.5$, where $p_2=20$dB, $p_1=5$dB, $\alpha=2$ and $\eta=1$.

Here the curve of $\hat R_s^+(d_0,d)$ illustrates the secrecy rate when Alice and Bob serve as transmitters in P1 and P2 respectively, which shows that the secrecy rate is low or zero when $d<0.2$  but the secrecy rate increases as $d$ increases.

The curve of $\bar R_s(d_0,d)$ illustrates the averaged secrecy rate resulting from two round-trip transmissions with opposite assignments for Alice and Bob, which shows that the averaged secrecy rate for any $0<d<1$ can be made no less than $\hat R_s(d_0,d)$ at $d=d_0=0.5$. In other words, the \emph{minimum} averaged secrecy rate $\min_d\bar R_s(d_0,d)$ in this case is
\begin{equation}\label{}
  \bar R_s(d_0,d)|_{d=d_0=0.5}=0.164
\end{equation}
  in bits per round-trip sample. This number is remarkable from a practical point of view as it could in principle allow One-Time-Pads (OTP) encryption in many practical situations. In practice, the channel between two nodes may not be fully loaded all the time, and hence during the quiescent periods the channel can be used for key generation via STEEP. Alternatively, if necessary for the highest security purpose, a fraction of total power and spectral resources of a classic channel can be easily reallocated for STEEP.

\section{Examples of Channel Coding for STEEP}

An advantage of STEEP is that it allows any established error correction code to be used by the P2 transmitter to encode $s_{2,k}$ so that the message encoded into $s_{2,k}$ can be decoded reliably by the P2 receiver but not by Eve.

Recall $\texttt{MSE}_U$ in \eqref{eq:sigma_s2|x1_y2} and $\texttt{MSE}_E$ in \eqref{eq:sigma_s2|z1_z2}.
With $c_1^2=0.3776$, $p_1=5$dB, $p_2=20$dB, $\alpha=2$, $\eta=1$ and $d=0.5$, we have $\texttt{MSE}_U=0.2046$
and $\texttt{MSE}_E = 0.2292$. Then
the \emph{effective} capacity in bits per sample of $s_{2,k}$ from Bob to Alice is $C_U=\log_2\frac{1}{\texttt{MSE}_U}=2.289$, and that from Bob to Eve is $C_E=\log_2\frac{1}{\texttt{MSE}_E}=2.125$.

%

The code rates for published LDPC codes (e.g., DVB-S2, 5G NR, WiFi, CCSDS) include
$R=1/4,1/3,2/5,1/2,3/5,2/3,3/4,4/5,5/6,8/9,9/10$.
Let $M$ (an integer power of 2) be the size of $M$-QAM symbol constellation for $s_{2,k}$. A desirable pair of $R$ and $M$ should be such that
\begin{equation}\label{eq:constraint}
  \frac{C_E}{R}<\log_2M<\frac{C_U}{R}.
\end{equation}
For $C_U=2.289$ and $C_E=2.125$, the above options of $R$ yield two desirable pairs of $R$ and $M$:  $(R,M)=(\frac{1}{4},2^9)$ and $(R,M)=(\frac{3}{4},2^3)$. Between the two choices, the first one provides more desirable margins around $\log_2 M$, i.e., the first pair yields $(\frac{C_E}{R},\frac{C_U}{R})=(8.5,9.156)$  while the second pair yields $(\frac{C_E}{R},\frac{C_U}{R})=(2.833,3.052)$.

%

Note that the effective capacity of the legitimate receiver (Alice or Bob) in P2 of STEEP  under the above choices of $c_1, p_1, p_2$ is invariant to $d$ (i.e., $2.289$). So the above code allows Alice or Bob to receive the packet associated with $s_{2,k}$ reliably regardless of $d$.

However, for $d<0.5$, the effective capacity at Eve is larger than $2.125$ (hence Eve may also receive the packet reliably) when Alice serves as the transmitter in P1, but is less than $2.125$ (hence Eve cannot receive the packet reliably) when Bob serves as the transmitter in P1.
A mirrored situation applies to $d>0.5$.

So, to establish a secret key of complete secrecy against Eve with any $0<d<1$, a pair of opposite sessions of STEEP (as described earlier) need to be conducted, and a privacy amplification as shown next needs to be followed.

\section{Privacy Amplification}

Using a standard (forward error correction) channel code  in P2 of each STEEP session, the information exchanged between the nodes in P2 is generally in partial secrecy against Eve, and the amount of secrecy depends on $d$. To generate a secret key with complete secrecy against Eve with $d$ unknown to users, privacy amplification is needed. Such a process involves the use of a (public) hash function \cite{Kate_2021} to compress the data already reliably exchanged between two nodes in P2 of each STEEP session. More specifically, let $\mathcal{D}_1$ denote the $N$-bits data  exchanged between the nodes in P2 of one STEEP session, and $\mathcal{D}_2$ be the $N$-bits data $\mathcal{D}_2$ exchanged between the nodes in P2 of an ``opposite'' STEEP session. Both of the STEEP sessions should be conducted consecutively so that $d$ remains the same (although unknown). The union set $\{\mathcal{D}_1,\mathcal{D}_2\}$ should be compressed by a hash function to a key $\mathcal{K}$ of size (number of bits) no larger than $2K\bar R_s(d_0,d)_{d=d_0=0.5}$. This key is now guaranteed to have the complete secrecy against Eve with any $d$.

Clearly, if Eve's position $d$ is known to users, then the privacy amplification process can be avoided by adopting a wiretap channel code in P2 of each STEEP session.

Unlike QKD, STEEP does not need a quantum channel, nor does it need back-and-forth iterative reconciliation iterations that would consume additional authenticated channel usages.

\section{Final Remarks}
This paper has shown useful properties of STEEP as an alternative to QKD. In many aspects, STEEP is much more competitive against QKD, which include the cost of deployment, the cost of required resources, the complexity of implementation, the compatibility with modern network architecture (which is virtually entirely based on classic or non-quantum channels), the secrecy rate (or equivalently secret-key rate after compression) needed for OTP encryption, and the robustness against constant eavesdropping. However, like QKD, STEEP (if implemented at the physical layer as shown in this paper) requires an analog channel between Alice and Bob assisted only by non-regenerative repeaters. Any regenerative relay or repeater could potentially leak all information to Eve. But modern optical repeaters for undersea cables are non-regenerative, and an overall end-to-end undersea cable can be modelled as Gaussian noise channel with path loss exponent no larger than two. This makes STEEP a much better alternative to QKD especially considering the current status of QKD. For developers of QKD, STEEP should definitely serve as an important benchmark in terms of cost and effectiveness.


\end{document}